\documentclass[a4paper,11pt]{article}
\usepackage{jheppub} 
\usepackage{amsmath, amssymb, amsthm, epsfig, fancyhdr,epsfig,multirow}
\usepackage[utf8]{inputenc}
\usepackage{amsmath}
\usepackage{amsfonts}
\usepackage{amssymb}
\usepackage{xcolor}
\usepackage{comment}
\usepackage{subcaption}
\usepackage[normalem]{ulem}
\usepackage{tabularx}
\usepackage{comment}
\usepackage{float}
\usepackage{graphicx}
\usepackage{appendix}
\usepackage[most]{tcolorbox}
\usepackage{physics}
\usepackage{mathtools}

\usepackage{tikz}
\usepackage{tikz-feynman}
\usepackage{braket}
\usepackage{cancel}
\usetikzlibrary{decorations.pathmorphing}

\title{Positivity at 1-Loop: Bounds on Photon and Gluon EFTs}

\author[a,1]{Jay Desai}
\author[a,2]{Diptimoy Ghosh}

\affiliation[a]{Department of Physics, Indian Institute of Science, Education and Research, Pune, India}

\emailAdd{desai.jay@students.iiserpune.ac.in}
\emailAdd{diptimoy.ghosh@iiserpune.ac.in}
\abstract{In this paper, we attempt to derive ``positivity" bounds on Photon and Gluon Effective Field Theories (EFTs) at one loop level. While for the Photon case, the one loop amplitude is IR finite and well defined in the forward limit, earlier studies failed to obtain a dispersive bound on dimension-12 operators due to the dependence of the ``arc integral" on the artificial low-energy scale.
We show that this awkward dependence can be taken care of by analysing the ultra-violet (UV) side of dispersion relation closely. In particular, we derive an IR safe and RG improved bound at 1-loop. Thereafter, we perform a similar analysis on the Gluon EFT, which has additional complications due to ill-defined forward limit and IR divergences at 1-loop. We show that even in this case, one can get a meaningful bound at 1-loop.}

\begin{document}
\maketitle
\flushbottom

\section{Introduction}
Effective Field Theories (EFTs) (see \cite{Skiba:2010xn,Manohar:2018aog} for a review) play a crucial role in understanding low energy physics. From a reductionist point of view, they can be obtained by integrating out heavy degree(s) of freedom from a full ``UV complete" theory. However, an effective field theorist can never have access to physics at the shortest scales. So, one writes down an EFT based on symmetries such as Lorentz Invariance, Gauge Invariance, etc. with arbitrary coefficients. The EFT is understood to be a low energy description of physics below a scale $\Lambda$. It is an expansion in $E/\Lambda$ (where $E$ is the relevant low energy scale of the process of interest), with higher dimensional operators suppressed by higher powers of $E/\Lambda$. So one can work at a given order by including only a finite number of operators in computations, and hence, only a finite number of coefficients, which can be measured by low energy experiments. Without the knowledge of the corresponding UV complete theory, these coefficients are a-priori arbitrary. One can ask if they can be constrained using general principles, without assuming any particular UV completion. In fact, in \cite{Adams:2006sv}, the authors showed that not all low-energy effective field theories described by a local, Lorentz invariant Lagrangian are compatible with a microscopic S-matrix satisfying the usual analyticity conditions and Unitarity. These constraints are known as \emph{positivity constraints} in the literature. \\\\
We first review the analysis of \cite{Adams:2006sv}, as relevant to this paper. Lorentz invariance, micro-causality and Unitarity of the $S$-matrix in the full UV theory and some (well-motivated) fundamental assumptions on the scattering amplitudes and their analytic structure, can be exploited to derive a relation between an IR contribution of an integral involving scattering amplitude (i.e. at low energies, where the EFT is valid) to a UV integral of its imaginary part, up to arbitrarily high energies. This relation between IR and UV parts of the amplitude is called a \emph{dispersion relation}. The UV side cannot be fully computed within the EFT but its positivity properties (which crucially relies on Unitarity of the full UV theory) can be exploited, which puts non-trivial constraints on the EFT coefficients of operators involved in the process, which a priori could have taken any values. \\ \\ 
One makes the following assumptions:\\ \begin{itemize}
    \item[\textbf{(i)}]  The \emph{forward} amplitude as a function of $s$ ($\mathcal{A}(s)$), in the complex $s$ plane, is analytic in the upper half plane, which follows from causality \cite{forward_analyticity_PhysRev.95.1612,Bros:1964iho, Lehmann} (see \cite{Eden, Sommer_Analyticity_Review} for a review and Sec 1.2 of \cite{Mizera:2023tfe} for a pedagogical explanation). The singularities are all on the Real-$s$ axis such as simple poles (due to tree-level exchanges) and branch cuts (due to logs at loop-level). 
\end{itemize}
That the amplitude is well defined at arbitrarily high energies is a consequence of Boost invariance. See \cite{SSBLI_Creminelli:2022onn, SSBLI_Creminelli:2023kze, SSBLI_Hui:2023pxc, SSBLI_Creminelli:2024lhd} for discussions on the fate of such bounds in theories without Lorentz Invariance.
\begin{itemize}
    \item [\textbf{(ii)}] The Froissart-Martin Bound ($\mathcal{A}(s)|_{s\rightarrow\infty}<\text{ const. } s\log^2(s)$) \cite{PhysRev.123.1053,Martin, Jin_PhysRev.135.B1375}, which was proved rigorously for only gapped theories, also holds for mass-less exchanges. 
    To make this more believable, we can use a mass regulator by assigning a mass $M$ to mass-less particles in the theory and at the end, take the limit $M\rightarrow 0$. We can make a slightly weaker assumption: $\lim_{s\rightarrow \infty}\mathcal{A}(s)/s^2=0$, known as ``Regge Bound" in the literature (see \cite{
    SharpBoundaries_Caron-Huot:2021rmr} for a discussion).\\
\end{itemize}
Consider a $2\rightarrow 2$ scattering process $$AB\rightarrow CD$$ 
with the Mandelstam variables defined as usual:
$$s=(p_A+p_B)^2, ~t=(p_C-p_A)^2, ~u=(p_D-p_A)^2$$
with $$s+t+u=m_A^2+m_B^2+m_C^2+m_D^2$$ 
where $A,B,C$ and $D$ are one-particle states with 4-momenta $p_A, p_B, p_C$ and $p_D$ respectively. Since $s,t$ and $u$ satisfy this constraint, a $2\rightarrow 2$ scattering amplitude $\mathcal{M}$ is a function of only two Mandelstam variables, usually chosen as $s$ and $t$. The scattering amplitude in the forward limit is denoted as
$$\lim_{t\rightarrow 0} \mathcal{M}(s,t)\equiv\mathcal{A}(s)$$
\\In this paper, since we are working with massless particles, the poles introduced at tree level (if any) will be all at origin. In case of massless exchanges, it is important to choose an appropriate $2\rightarrow 2$ process such that the $t\rightarrow 0$ limit is well defined.
\begin{center}
\begin{tikzpicture}
\def\gap{0.2}
\def\R{3}
\def\r{0.45}

\draw [line width=0.05mm, gray ,<->] (-1.26*\R, 0) -- (1.26*\R,0) node [right]{};
\draw[-, decorate, decoration={zigzag, segment length=6, amplitude=1.5}, blue] (-1.2*\R, 0) -- (1.2*\R, 0) node [above left, black] {};

\draw [help lines,<->] (0, -\gap) -- (\r,-\gap);
\draw [help lines,<->] (-1.26*\R, 0) -- (1.26*\R,0);
\draw [help lines,<->] (0, -0.5*\R) -- (0, 1.25*\R);
\draw[line width=1pt,   decoration={ markings,
  mark=at position 0.2455 with {\arrow[line width=1.2pt]{<}},  
  mark=at position 0.765 with {\arrow[line width=1.2pt]{<}},
  mark=at position 0.93 with {\arrow[line width=1.2pt]{<}},
  mark=at position 0.7 with {\arrow[line width=1.2pt]{<}}},
  postaction={decorate}]
  let
     \n1 = {asin(\gap/2/\R)},
     \n2 = {asin(\gap/2/\r)}
  in (\n1:\R) arc (\n1:180-\n1:\R)
  -- (-\n2:-\r) arc (\n2:-180-\n2:-\r) -- (\n1:\R);

\node at (\gap/2+0.1,-2*\gap){\small$s_0$};
\node at (3.6,-0.37){$Re(s)$};
\node at (0.66,3.53) {$Im(s)$};
\node at (-0.7,0.55) {$\cap_{s_0}$};
\node at (-1.8,2.8) {$\cap_{\infty}$};

\end{tikzpicture}
\end{center}
Consider the contour as shown above, as was first considered in \cite{Bellazzini:2020cot}. Since this does not enclose any residues, we have
$$ \oint  ds \frac{\mathcal{A}(s)}{s^{2n+1}}=0$$
for some positive integer $n$.
We can decompose this as follows:
$$\oint=\int_{\infty}^{s_0}+\int_{\cap_{s_0}}^{~}+\int_{-s_0}^{-\infty}+\int_{\cap_\infty}$$
where second and fourth integrals are over semi-circular arcs of radius $s_0$ and $\infty$ respectively.
After a substitution $s\rightarrow -s$ in the third integral, it can be clubbed with the first.
$$0=\int_{s_0}^{\infty} ds ~\frac{\mathcal{A}(-s+i\epsilon)-\mathcal{A}(s+i\epsilon)}{s^{2n+1}}+\int_{\cap_{s_0}}ds~\frac{\mathcal{A}(s)}{s^{2n+1}}+\int_{\cap_\infty}ds~\frac{\mathcal{A}(s)}{s^{2n+1}}$$
The last term on the RHS - the integral over the arc at infinity - gives zero (for $n>0$) due to Regge boundedness. Then, we have
$$\int_{\cap_{s_0}}ds~\frac{\mathcal{A}(s)}{s^{2n+1}}=\int_{s_0}^\infty ds ~\frac{\mathcal{A}(s+i\epsilon)-\mathcal{A}(-s+i\epsilon)}{s^{2n+1}}$$
With analytic continuation (or real analyticity: 
$\mathcal{A}(s)=\mathcal{A}^*(s^*)$) and choosing an $s$-$u$ crossing-symmetric amplitude \cite{crossing, Trueman:1964zzb, crossing_spin}, ($\mathcal{A}(s)=\mathcal{A}(-s)$), we have for $s$ on real axis: $\mathcal{A}(s+i\epsilon)=\mathcal{A}^*(s-i\epsilon)=\mathcal{A}^*(-s+i\epsilon)$, (which we refer to as ``crossing symmetry") so that we can relate the negative $s$ branch to the (physical) positive one. Under this assumption, we get 
\begin{equation}
    \int_{\cap_{s_0}}\frac{ds}{i\pi}~\frac{\mathcal{A}(s)}{s^{2n+1}}=\frac{2}{\pi}\int_{s_0}^\infty ds ~\frac{Im\mathcal{A}(s)}{s^{2n+1}}
\end{equation}
We make use of the Optical Theorem, which for massless particles in the initial state reads: $Im\mathcal{A}(s)=s\sigma(s)$. So, we finally get the dispersion relation
\begin{equation}\label{PA}
    \int_{\cap_{s_0}}\frac{ds}{i\pi}~\frac{\mathcal{A}(s)}{s^{2n+1}}=\frac{2}{\pi}\int_{s_0}^\infty ds ~\frac{\sigma(s)}{s^{2n}}
\end{equation}
One can take $s_0<\Lambda^2$, so that the LHS can be computed within the EFT, in terms of Wilson coefficients of EFT operators ($\mathcal{A}=\mathcal{A}_{\text{EFT}}$ for $s<\Lambda^2$). 
The integer $n$ has to be chosen to get the desired contribution of Wilson coefficients from the LHS, which we can then constrain to be positive due to the RHS being positive.
\begin{equation} \label{positivity}
    \textit{Amplitude's Positivity:}~~~\int_{\cap_{s_0}}\frac{ds}{i\pi}~\frac{\mathcal{A}_{\text{EFT}}(s)}{s^{2n+1}}>0
\end{equation}
For instance, consider a tree level contribution 
$$\mathcal{A}\sim c_{_2}\frac{s^2}{\Lambda^4}+c_{_6}\frac{s^6}{\Lambda^{12}}+...$$
Then, $n=1$ in (\ref{positivity}) gives $c_{_2}>0$ and $n=3$ gives $c_{_6}>0$. \\ \\ 
This analysis has been explicitly done at tree level in various contexts (e.g. \cite{Adams:2006sv,Bellazzini_PB,TreePB_Creminelli:2024lhd,Azatov:2021ygj,Ghosh:2022qqq,PA_EliasMiro:2022xaa,PA_Low:2009di,PA_Tolley:2020gtv,PA_Zhang:2018shp, SumRules_Bellazzini:2014waa,PA_Nicolis:2009qm,Chakraborty:2024ciu}) in the forward limit and \cite{NonForward_Loop_Vecchi:2007na,PA_Manohar:2008tc, NonForward_Nicolis:2009qm,deRham:2017_ScalarPA, deRham:2017_UVComplete, Arkani-Hamed_EFTHedron,TreeNF_Berman:2023jys, ExtremalEFTs_Caron-Huot:2020cmc,SharpBoundaries_Caron-Huot:2021rmr, Vichi_LbyL_bounds,treePA_nf_Sinha:2020win,TreeNF_Bertucci:2024qzt} in the non-forward limit. To obtain tree level bounds on higher dimensional operators, one needs the extra assumption that the UV theory is weakly coupled via some additional small parameter to separate tree and loop contributions at the same order in $p/\Lambda$.\\ \\
In this paper, our focus would be to investigate forward limit dispersive positivity bounds beyond tree level. As one can expect, there are some immediate difficulties encountered at 1-loop such as IR divergence and $\log(t)$ divergence in the forward limit. Attempts to include loop contributions have been made \cite{Bellazzini:2020cot,EFTHedron_loop_Peng:2025klv,BPS,IRside, FiniteMpl_Beadle:2025cdx, Bellazzini_loops, Li:2022aby}, etc. See also \cite{UVIRNonForward_Beadle:2024hqg,GravityLoops_Caron-Huot:2024tsk,GravityLoops_Arkani-Hamed:2021ajd}, and the references therein, in context of gravity. In particular, \cite{BPS} attempted this in case of EFT of Photons, which although free of IR and $\log(t)$ divergences, got a 1-loop correction dependent on $\log(s_0)$, disallowing the $s_0\rightarrow 0$ limit to drop the contributions from higher derivative operators. We revisit this problem and show that we can get rid of this $\log(s_0)$ dependence, and get a meaningful dispersive bound on dimension-12 operators. \\ \\
We also attempt to derive positivity bounds at 1-loop level for EFT of Gluons (extending the tree level bounds in \cite{Ghosh:2022qqq}) where the situation is way more complicated due to IR and $\log t$ divergences at 1-loop. We show that even in this case, these difficulties can be removed.
\\ \\
The rest of the paper is organized as follows: In Sec \ref{gen}, we introduce the notations and some definitions. In Sec \ref{EH12}, we derive $s_0$-independent 1-loop corrections to Positivity Bounds on dim-12 operators in EFT of Photons. In Sec \ref{QCDEH}, we derive 1-loop corrections to bounds on dim-8 operators in EFT of Gluons. \\\\

\section{Notations and Conventions}\label{gen}
We use the mostly-minus signature for Minkowski metric: $$\eta_{\mu\nu}=\text{diag}(1,-1,-1,-1)$$
We compute all amplitudes in the Centre of Mass frame $(\Vec p_1+\Vec p_2=\Vec p_3+\Vec p_4=0)$, with 4-momenta $p_1, p_2$ incoming and $p_3, p_4$ outgoing.

 $$p_1 = (E,0,0,E) \text{,~~} p_2=(E,0,0,-E)$$ 
 $$p_3 = (E,E\sin\theta,0,E\cos\theta) \text{,~~} p_4=(E,-E\sin\theta,0,-E\cos\theta)$$
 Polarizations corresponding to helicity $(\lambda=\pm 1)$ eigen-states are (see Appendix B.2 of \cite{BPS} for a derivation from group theory perspective)
 $$\epsilon^{\pm}(p_1)=\pm\frac{1}{\sqrt{2}}(0,\mp 1,-i,0) \text{,~~} \epsilon^{\pm}(p_2)=\pm\frac{1}{\sqrt{2}}(0,\pm 1,-i,0)  $$
 $$\epsilon^{\pm}(p_3)=\pm\frac{1}{\sqrt{2}}(0,\mp \cos\theta,-i,\pm \sin\theta) \text{,~~} \epsilon^{\pm}(p_4)=\pm\frac{1}{\sqrt{2}}(0,\pm \cos\theta,-i,\mp \sin\theta)  $$ \\
We take the initial and final state to be in the following superposition of helicity eigen-states.
$$\ket{\psi}_\pm \sim \frac{\ket{++}+\ket{+-}+\ket{-+}\pm\ket{--}}{2}$$
The forward amplitude corresponding to this state is
\begin{equation}\label{csamp}
\mathcal{A}^\pm=\bra{\psi}T\ket{\psi}_\pm=\frac{\mathcal{A}_{++++}+\mathcal{A}_{+-+-}\pm\mathcal{A}_{++--}}{2} 
\end{equation}
It was shown in \cite{BPS} that for photons, this amplitude
is crossing symmetric, using crossing relations first discussed in \cite{Trueman:1964zzb} and more recently in \cite{Hebbar_Spinning}. Here, $$\mathcal{A}_{\lambda_1\lambda_2\lambda_3\lambda_4}\equiv\bra{p_1,\lambda_3;p_2,\lambda_4}T\ket{p_1,\lambda_1;p_2,\lambda_2}$$
denotes the forward scattering amplitude, but with helicities $\lambda_1\lambda_2\rightarrow\lambda_3\lambda_4$. \\ \\
For computing the amplitudes, we take help of the Mathematica packages: FeynCalc \cite{feyncalc}, FeynArts \cite{feynarts}, FeynRules \cite{feynrules} and FeynHelpers \cite{feynhelpers}. \\\\ Note: FeynCalc uses the definition $$\varepsilon=\frac{4-D}{2}$$ which we convert to the more convenient $\epsilon=4-D$ i.e. 
$$\varepsilon_{F.C.}=\frac{\epsilon}{2}$$
The gluon field strength tensor and its dual are defined as
$$G^{a}_{\mu\nu}=\partial_\mu A^{a}_\nu-\partial_\nu A_\mu^a+g  f^{a}_{~bc}A^b_\mu A_\nu^c~\text{  and  }~\widetilde{G}^{a}_{\mu\nu}=\frac{1}{2}\epsilon_{\mu\nu\rho\sigma}G^{a\;\rho\sigma}\;\; (\epsilon_{0123}=+1)$$
respectively. Here, $A_\mu =t_{a}A_\mu^{a}$ is the gluon field, $t_a$ are the generators of SU(3); $a$, $b$, $c$ = 1, 2, ..., 8 are colour indices; $g$ is the coupling constant of the strong force; $f^{abc}$ and $d^{abc}$ are the totally anti-symmetric and symmetric structure constants of SU(3) respectively, defined as
$$\left[t^a,t^b\right]=if^{ab}_{~~c}t^c,~~ \left\{t^a, t^b\right\}=\frac{1}{3} \delta^{a b}+d^{a b}_{~~ c} t^c$$
\section{EFT of photons}\label{EH12}
\begin{figure}[h!]
    \centering
    \includegraphics[width=0.5\linewidth]{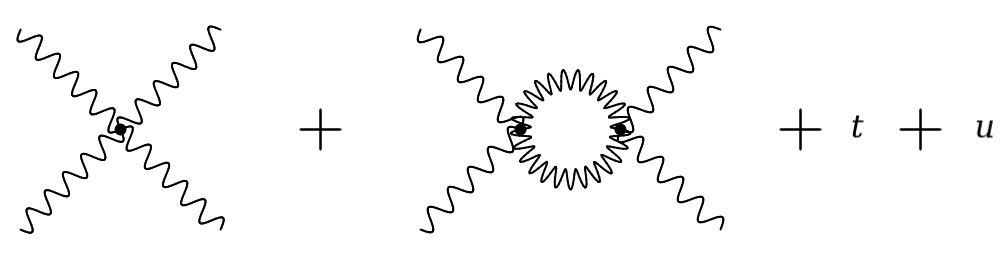}
    \caption{$\gamma\gamma\rightarrow\gamma\gamma$ up to $\mathcal{O}(s^4/\Lambda^8)$}
    \label{fig:dim-8+12} 
\end{figure}

We begin by considering EFT of only photons, with operators of mass dimension 8, 10 and 12 that contribute to $2\rightarrow 2$ scattering, listed in \cite{BPS}. 
\begin{eqnarray}
    \mathcal{L} & = & -\frac{1}{4}F_{\mu\nu}F^{\mu\nu}+\frac{c_{_1}}{\Lambda^4}(F_{\mu\nu}F^{\mu\nu})^2+\frac{c_{_2}}{\Lambda^4}F_{\mu\nu}F^{\nu\sigma}F_{\sigma\rho}F^{\rho\mu} + \frac{c_{_3}}{\Lambda^6}F_{\alpha\beta}\partial^\beta F_{\mu\nu}\partial^\alpha F^{\nu\rho} F_\rho^{~\mu}\nonumber\\
    & &  + \frac{c_{_4}}{\Lambda^6}F_{\alpha\beta}F^{\alpha\beta}\partial_\mu F_{\rho\sigma}\partial^\mu F^{\rho\sigma}+\frac{c_{_5}}{\Lambda^6}\partial_\alpha F_{\mu\nu}\partial^\alpha F_{\rho\sigma}F^{\nu\rho}F^{\sigma\mu}+\frac{c_{_6}}{\Lambda^8}\partial_\nu F_{\alpha \rho} \partial^\alpha F^{\nu \sigma} \partial^\delta F^{\beta \rho} \partial_\beta F_{\delta \sigma}\nonumber\\
    & & +
    \frac{c_{_7}}{\Lambda^8} F_{\rho \beta} \partial_\alpha F_{\sigma \gamma} \partial^{\beta} F^{\sigma \delta} \partial_{\delta} \partial^{\gamma} F^{\rho \alpha}
    +\frac{c_{_8}}{\Lambda^8} F_{\alpha \beta} F_{\sigma \gamma} \partial^\alpha \partial_{\rho} F^{\gamma \delta} \partial^\beta \partial_{\delta} F^{\rho \sigma} \label{LEH}
\end{eqnarray}
The dim-10 operators contribute as $\sim (s^3+u^3)$ in the tree level $s$-$u$ symmetric amplitude, which vanishes in the forward limit and hence, $c_{_{3,4,5}}$ cannot be bounded by \emph{this} method at tree level. In fact, in Sec 4.4 of \cite{BPS}, they show that dim-10 coefficients cannot be bounded analytically. It would require two insertions of dim-10 operators to be able to derive some bounds on them, which at tree level can only come from exchange diagrams, which require a 3-point vertex which does not exist in this case. So we set $c_{_3}=c_{_4}=c_{_5}=0$ as they are not relevant for us. Also, we want to study loop-effects and the first loop contribution comes from two insertions of dim-8 operators at order $s^4/\Lambda^8$, and the first tree level contribution to this order comes from one dim-12 insertion in a contact diagram (see Fig. \ref{fig:dim-8+12}). \\ \\
The tree level bounds in this theory were first obtained by \cite{Vichi_LbyL_bounds} and the loop corrections were first studied by \cite{BPS}. As pointed out in the latter, including the 1-loop contribution (from two dim-8 insertions) leads to a $\log(s)$ dependence in the corrected bounds. We show that taking into account (i) the UV side of the dispersion relation in the EFT regime and (ii) the precise running of the dim-12 coefficients, leads to a meaningful 1-loop corrected bound.
\\ \\
We consider the $s$-$u$ symmetrized amplitude, as defined in Eq (\ref{csamp}):
$$\mathcal{A}^\pm=\frac{1}{2}\left(\mathcal{A}^{++++}+\mathcal{A}^{+-+-}\pm\mathcal{A}^{++--}\right)=\mathcal{A}^\pm_{\text{tree}}+\mathcal{A}^\pm_{\text{uv}}+\mathcal{A}^\pm_{\text{finite}}$$
where we have separated the tree amplitude, UV divergence from the loop and the finite part from the loop.
In Section \ref{fullcomp}, we show the full computation which inevitably has a lot of notation, making the physics and main conclusions obscure. So in the next section, we do the analysis with only $c_{_2}, c_{_7}\neq 0$ to make the main ideas transparent. Section \ref{fullcomp} will just be a generalization. 
\subsection{Simpler Case}\label{simplercase}
In this section, we derive the bounds on the theory with just one dim-8 and one dim-12 operator. The main ideas and conclusions are made clear in this sub-section, whereas the next sub-section (Sec \ref{fullcomp}) only contains more detailed calculations, generalizing this sub-section.
\begin{equation*}
    \mathcal{L} \supset -\frac{1}{4}F_{\mu\nu}F^{\mu\nu}+\frac{c_{_2}}{\Lambda^4}F_{\mu\nu}F^{\nu\sigma}F_{\sigma\rho}F^{\rho\mu}  +
    \frac{c_{_7}}{\Lambda^8} F_{\rho \beta} \partial_\alpha F_{\sigma \gamma} \partial^{\beta} F^{\sigma \delta} \partial_{\delta} \partial^{\gamma} F^{\rho \alpha}
\end{equation*}
The $2\rightarrow 2$ forward scattering amplitudes up-to $\mathcal{O}(s^4/\Lambda^8)$ are\footnote{We have redefined $$\frac{2}{\epsilon}+\log(4\pi e^{-\gamma_E}) \rightarrow \frac{2}{\epsilon}$$}

\begin{eqnarray}
    \mathcal{A}^{++++} &=& \frac{6c_{_2}}{\Lambda^4}s^2+\frac{(c_{_7}/2)}{\Lambda^8}s^4+\frac{77}{40\pi^2}\frac{c_{_2}^2}{\Lambda^8}s^4\left(\frac{2}{\epsilon}+\log(\mu^2/s)\right)+\frac{c_{_2}^2}{\Lambda^8} s^4\frac{1770i\pi+4489}{1200\pi^2} \nonumber\\
    \mathcal{A}^{+-+-} &=& \frac{6c_{_2}}{\Lambda^4}s^2+\frac{(c_{_7}/2)}{\Lambda^8}s^4+\frac{77}{40\pi^2}\frac{c_{_2}^2}{\Lambda^8}s^4\left(\frac{2}{\epsilon}+\log(\mu^2/s)\right)+\frac{c_{_2}^2 }{\Lambda^8}s^4\frac{540i\pi+4489}{1200\pi^2}\nonumber \\  
    \mathcal{A}^{++--} &=& \frac{4c_{_2}}{\Lambda^4}s^2-{0}\frac{s^4}{\Lambda^8}+\frac{5}{2\pi^2}\frac{c_{_2}^2}{\Lambda^8}s^4\left(\frac{2}{\epsilon}+\log(\mu^2/s)\right)+\frac{c_{_2}^2}{\Lambda^8}s^4\frac{50i\pi+197}{40\pi^2} 
\end{eqnarray}
and the crossing symmetrized amplitude as defined in Eq (\ref{csamp}) (with `$-$' sign)
$$\mathcal{A}=\frac{\mathcal{A}_{++++}+\mathcal{A}_{+-+-}-\mathcal{A}_{++--}}{2} $$
is\footnote{We take the branch where $\log(-s)=\log(s)+i\pi$. So the combination $2\log(s)-i\pi$ has the correct crossing symmetry.}
\begin{equation}\mathcal{A}=4c_{_2}\frac{s^2}{\Lambda^4}+{(c_{_7}/2)}\frac{s^4}{\Lambda^8}+\frac{27c_{_2}^2}{40\pi^2}\frac{s^4}{\Lambda^8}\left(\frac{2}{\epsilon}+\log(\mu^2/s)+i\frac{\pi}{2}\right)+\frac{767c_{_2}^2}{600\pi^2}\frac{s^4}{\Lambda^8}\end{equation}
We will use this amplitude for the dispersive analysis as reviewed in the Introduction. The UV divergence will have a non-zero contribution to the arc-integral. So we also need to add a counter term to renormalize it i.e. 
$$\frac{\delta c_{_7}}{\Lambda^8} F_{\rho \beta} \partial_\alpha F_{\sigma \gamma} \partial^{\beta} F^{\sigma \delta} \partial_{\delta} \partial^{\gamma} F^{\rho \alpha}$$
to cancel the $1/\epsilon$ UV divergence at order $s^4/\Lambda^8$:
\begin{equation}\label{ct7}
    \frac{\delta c_{_7}}{2}=-\frac{27c_{_2}^2}{40\pi^2}~\frac{2}{\epsilon}+\frac{f_{_7}}{2}
\end{equation}        
where $f_{_7}$, an arbitrary finite piece, decides the Renormalization Scheme and defines the coupling. Of course, this only cancels the divergence in the crossing symmetric amplitude; the more detailed calculation requires contributions from $\delta c_{_{6,8}}$, shown in Sec \ref{fullcomp}. After adding the counter term,
the Amplitude is
\begin{equation}\mathcal{A}=4c_{_2}\frac{s^2}{\Lambda^4}+\frac{c_{_7}+f_{_7}}{2}\frac{s^4}{\Lambda^8}+\frac{27c_{_2}^2}{40\pi^2}\frac{s^4}{\Lambda^8}\left(\frac{2}{\epsilon}+\log(\mu^2/s)+i\frac{\pi}{2}\right)+\frac{767c_{_2}^2}{600\pi^2}\frac{s^4}{\Lambda^8}\end{equation}
Using
$$\int_{\cap_{s_0}}\frac{ds'}{i\pi}\frac{1}{s'}(A(2\log( s'/\mu^2)-i\pi) + B)= 2A\log (s_0/\mu^2) + B $$
the relevant arc integral (with $n=2$) gives
$$\int_{\cap_{s_0}}\frac{ds'}{i\pi}\frac{\mathcal{A}(s')}{s'^5}=0+\frac{c_{_7}+f_{_7}}{2\Lambda^8}-\frac{27}{40\pi^2}\frac{c_{_2}^2}{\Lambda^8}\log (s_0/\mu^2)+\frac{767}{600\pi^2}\frac{c_{_2}^2}{\Lambda^8} $$
One can choose the scheme\footnote{With this scheme, the $s^4$ contribution to the amplitude comes only from dim-12.}
\begin{equation}\label{schemechoice}
    \frac{f_{_7}}{2}=-\frac{767c_{_2}^2}{600\pi^2}
\end{equation}
for simplicity. This corresponds to the scheme chosen in \cite{BPS} (see Eq (D.13) of their arxiv version [v1]) and reproduces (the simplified version of) their Positivity Bound (Eq (1.27)):
\begin{equation}
    c_{_7}(\mu)-\frac{27c_{_2}^2}{20\pi^2}\log(s_0/\mu^2)>0
\end{equation}
As they note, this bound is not meaningful for small values of $s_0$ as one would like. We show that the $\log s_0$ dependence cancels, and one can safely take $s_0\rightarrow 0$ to minimize the corrections due to higher dimensional operators (of $\mathcal{O}(s_0/\Lambda^2)$) and derive an exact meaningful bound (as explained after Eq (\ref{simple bound})). \\ \\
On the right hand side of the dispersion relation, Eq (\ref{PA}), the integral can be separated into IR and UV contributions: 
\begin{equation}\label{split}
    \int_{\cap_{s_0}}\frac{ds}{i\pi}~\frac{\mathcal{A}(s)}{s^5}=\frac{2}{\pi}\int_{s_0}^{{\Lambda}^2} ds \frac{\sigma(s)}{s^4}+\frac{2}{\pi}\int_{{\Lambda}^2}^{\infty} ds \frac{\sigma(s)}{s^4}
\end{equation}
 From Optical Theorem, we have the cross-section (to order $s^3/\Lambda^8$):
$$\sigma(s)=Im\mathcal{A}(s)/s=\frac{27}{80\pi}\frac{c_{_2}^2}{\Lambda^8}s^3 ~~~(\text{for}~ s<\Lambda^2)$$
Using this, we can include the contribution in the EFT regime from RHS i.e. the first term from RHS in Eq (\ref{split})\footnote{To be precise, the upper limit of Eq (\ref{split}) may be thought of as slightly less than $\Lambda^2$. We keep it as $\Lambda^2$ for simplicity as this will not affect our conclusions.}:
\begin{equation}\label{IPA simp}
    \implies\frac{c_{_7}}{2\Lambda^8}-\frac{27}{40\pi^2}\frac{c_{_2}^2}{\Lambda^8}\log (s_0/\mu^2)=-\frac{2}{\pi}\frac{27c_{_2}^2}{80\pi\Lambda^8}\log(s_0/\Lambda^2)+\text{positive}
\end{equation}
The $s_0$ dependence cancels from second term on LHS and first term on RHS; so, we get an $s_0$ independent bound:
\begin{equation}\label{simple bound}
    c_{_7}(\mu)-\frac{27}{20\pi^2}c_{_2}^2\log({\Lambda}^2/\mu^2)>0
\end{equation} 
There will be $\mathcal{O}(s_0/\Lambda^{10})$ corrections  (from higher dimensional operators) on the LHS of Eq (\ref{IPA simp}), which become $\mathcal{O}(s_0/\Lambda^2)$ in the bound Eq (\ref{simple bound}), and can be dropped by taking $s_0\rightarrow 0$. The corrections introduced in the cross section on the RHS, although could be $\mathcal{O}(1)$ (since the integral is up to $\Lambda^2$), are all positive (since Optical Theorem is true order-by-order in $p/\Lambda$), and can be clubbed with the term `positive' in Eq (\ref{IPA simp}), and the final bound obtained is actually exact (no $p/\Lambda$ suppressions). \\ \\
The $\log$-term in Eq (\ref{simple bound}) comes from the RG running of $c_{_7}(\mu)$ (which we derive below: Eq (\ref{c7running})). The RG calculation details can be skipped by jumping to Eq (\ref{c7running}) (where we have restored the scheme dependence).
\subsubsection*{RG running}
Now, let's compute the RG running of $c_{_7}(\mu)$ to rewrite this bound in terms of the UV scale $\Lambda$, which is the scale at which EFT matching minimizes the $\log$s. First, we write the Lagrangian in terms of the renormalized couplings and fields:
$$A_\mu \rightarrow \sqrt{Z_A}A_\mu,~ c_{_2} \rightarrow Z_2 c_{_2},~ c_{_7} \rightarrow Z_7 c_{_7}$$
where $$Z_A\equiv 1+\delta_A, ~Z_2\equiv 1+\delta_2, ~Z_7\equiv 1+\delta_7$$ \\
We will have combinations of the type $$Z_A^2Z_i=(1+\delta_A)^2(1+\delta_i)= 1 + (2\delta_A+\delta_i)+\mathcal{O}(\delta^2)$$
In terms of these,
\begin{eqnarray*}
    \mathcal{L} & = & -\frac{1}{4}F_{\mu\nu}F^{\mu\nu}+\frac{c_{_2}}{\Lambda^4}F_{\mu\nu}F^{\nu\sigma}F_{\sigma\rho}F^{\rho\mu} +
    \frac{c_{_7}}{\Lambda^8} F_{\rho \beta} \partial_\alpha F_{\sigma \gamma} \partial^{\beta} F^{\sigma \delta} \partial_{\delta} \partial^{\gamma} F^{\rho \alpha}\\
    & & -\delta_A\frac{1}{4}F_{\mu\nu}F^{\mu\nu}+(2\delta_A+\delta_2)\frac{c_{_2}}{\Lambda^4}F_{\mu\nu}F^{\nu\sigma}F_{\sigma\rho}F^{\rho\mu}  +(2\delta_A+\delta_7)\frac{c_{_7}}{\Lambda^8} F_{\rho \beta} \partial_\alpha F_{\sigma \gamma} \partial^{\beta} F^{\sigma \delta} \partial_{\delta} \partial^{\gamma} F^{\rho \alpha}\\
\end{eqnarray*}
So, the counter term we computed in Eq (\ref{ct7}) is
$$\delta c_{_7} = (2\delta_A + \delta_7)c_{_7}$$ 
We have neglected $\delta_A^2$ terms as they are $\mathcal{O}(p^8/\Lambda^{8})$ and contribute to $\delta c_{_2}/\Lambda^4$ at  $\mathcal{O}(p^{12}/\Lambda^{12})$. Since we are working at $\mathcal{O}(p^8/\Lambda^8)$, we need $\delta_A$ at $\mathcal{O}(p^4/\Lambda^4)$. At this order, it only receives contribution from a self-energy loop, which is scaleless and hence vanishes in dim-reg. So we can take $\delta_A=0$ in our computations\footnote{In fact, at the order $p^8/\Lambda^8$, the correction to photon propagator comes from a 2-loop diagram, which from dimensional analysis and power counting arguments is expected to be of the form $$\frac{g_{\mu\nu}-\frac{p_\mu p_\nu}{p^2}}{p^2}\frac{p^8}{\Lambda^8}c_ic_j\log^2(\mu^2/p^2)\left(\frac{a}{\epsilon^2}+\frac{b}{\epsilon}\right)$$
So, the photon propagator $$\big<A_\mu A_\nu \big>\sim\left(g_{\mu\nu}-\frac{p_\mu p_\nu}{p^2}\right)\frac{1}{p^2(1+\left(\frac{a}{\epsilon^2}+\frac{b}{\epsilon}\right)c_i c_j\frac{p^8}{\Lambda^8}\log^2(\mu^2/p^2)-\delta_A)}$$ has a finite residue at $p^2=0$ pole if $\delta_A=0$. It seems that this argument can be generalized to show that $\delta_A=0$ to all orders in $p/\Lambda$, since more EFT insertions will only lead to more powers of $p^2/\Lambda^2$ in the corrections, which vanish at $p^2=0$. This is possible because of our EFT being massless, so there is no additional mass scale to give $M^8/\Lambda^8$ type contributions. \label{WFR}} 
and
$$\delta c_{_7}=\delta_7c_{_7}$$
Then the running of $c_{_7}$ can be computed from the RGE $$\mu\frac{d}{d\mu}(\mu^\epsilon Z_7 c_{_7})=\mu\frac{d}{d\mu}(\mu^\epsilon (c_{_7} + \delta c_{_7}))=0$$ 
$$\implies \epsilon (c_{_7}+\delta c_{_7})+\mu\frac{d}{d\mu}(c_{_7}+\delta c_{_7})=0$$
Since $\delta c_{_7}\sim c_{_2}^2/\epsilon$, we need the running of $c_{_2}$.
There can't be a dim-8 counter term $\frac{\delta c_{{\text{dim-}8}}}{\Lambda^4}F^4$ since this would require a loop with two insertions of dim-6 operators $\frac{c_{\text{dim-}6}}{\Lambda^2}\mathcal{O}_{\text{dim-6}}$ which don't exist in our case. So the $\beta$ function of a dim-8 coefficient $c_{_{2}}$ is obtained by 
$$\mu\frac{d}{d\mu}(\mu^\epsilon c_{_{2}})=0$$
\begin{equation}\label{betac2}
\implies\mu\frac{d}{d\mu}c_{_{2}}=-\epsilon c_{_{2}}
\end{equation}
i.e. $c_{_{2}}$ (or any other dim-8 coefficient) doesn't run.
So, 
\begin{eqnarray}
    \mu\frac{d}{d\mu}\delta c_{_7}&=&\mu\frac{d}{d\mu}\left(-\frac{27c_{_2}^2}{20\pi^2}\frac{2}{\epsilon}+f_{_7}\right) \nonumber\\
    &=&-\frac{27}{5\pi^2\epsilon}c_{_2}\mu\frac{d}{d\mu}c_{_2}+\mu\frac{d}{d\mu}f_{_7} \label{int2}\\
    &=&+\frac{27c_{_2}^2}{5\pi^2}+\mu\frac{d}{d\mu}f_{_7}+\mathcal{O}(\epsilon) \nonumber
\end{eqnarray}
Note that in the step (\ref{int2}), we need the $\beta$ function of $c_{_2}$ to $\mathcal{O}(\epsilon)$ (Eq (\ref{betac2})) because of the form of UV divergence $\sim c_{_2}^2/\epsilon$.
$$\therefore \epsilon (c_{_7}+\delta c_{_7})+\mu\frac{d}{d\mu}(c_{_7}+f_{_7})+\frac{27c_{_2}^2}{5\pi^2}=0$$
Setting terms of $\mathcal{O}(\epsilon)=0$, we get
$$\mu\frac{d}{d\mu}(c_{_7}+f_{_7})(\mu)=-\frac{27c_{_2}^2}{10\pi^2}$$
Integrating for $C_{_7}\equiv c_{_7}+f_{_7}$, we get
$$C_{_7}(\mu)-C_{_7}(\mu_1)=-\frac{27c_{_2}^2}{10\pi^2}\log(\mu/\mu_1)$$
or
\begin{equation}\label{c7running}
    C_{_7}(\mu)=C_{_7}(\Lambda)+\frac{27c_{_2}^2}{10\pi^2}\log(\Lambda/\mu)
\end{equation}
where we have set $\mu_1=\Lambda$, where the EFT is matched to the UV theory.
Then, the bound on $c_{_7}$ (Eq (\ref{simple bound})) can be written as
$$c_{_7}(\Lambda)+f_{_7}(\Lambda)+\frac{27c_{_2}^2}{10\pi^2}\log(\Lambda/\mu)-\frac{27c_{_2}^2}{20\pi^2}\log(\Lambda^2/\mu^2)+\frac{767c_{_2}^2}{300\pi^2}>0$$
The third and fourth terms cancel, and with the scheme choice Eq (\ref{schemechoice})\footnote{Since $c_{_2}$ doesn't run, this scheme choice is preserved at any energy scale.}, we have the bound at the scale $\Lambda$:
\begin{equation}\label{finalboundsimp}
    c_{_7}(\Lambda)>0
\end{equation} 
In Appendix \ref{comment}, we give some comments about this result and the bound at lower scales.\\\\
One can also substitute the running in the Amplitude directly:
$$\mathcal{A}(s)=\frac{4c_{_2}}{\Lambda^4}s^2+\frac{(c_{_7}+f_{_7})(\Lambda)}{2}\frac{s^4}{\Lambda^8}-\frac{27}{80\pi^2}\frac{c_{_2}^2}{\Lambda^8}s^4\left(2\log(s/\Lambda^2)-i\pi\right)+\frac{767}{600\pi^2}\frac{c_{_2}^2}{\Lambda^8}s^4$$
and perform the analysis on this to again get the bound (\ref{finalboundsimp}).\\\\
So, we conclude that the $\log s_0$ dependence can be cancelled by taking into account the RHS of dispersion relation in the EFT regime and the tree level bound $c_{_7}>0$ is modified at 1-loop only by RG effects.

\subsection{Full Computation}\label{fullcomp}
In this subsection, we generalize the procedure followed in Sec \ref{simplercase} with all dim-8 and dim-12 operators present (Eq (\ref{LEH})). The main ideas and concepts remain the same as in the previous subsection.
\subsubsection*{Tree}
The tree level amplitude has contributions from both dim-8 and dim-12 operator insertions.
$$\mathcal{A}^{++++,+-+-}_{\text{tree}}=2(4c_{_1}+3c_{_2})\frac{s^2}{\Lambda^4}+{\frac{2c_{_7}-3c_{_6}}{4}}\frac{s^4}{\Lambda^8}$$
$$\mathcal{A}^{++--}_{\text{tree}}=4(4c_{_1}+c_{_2})\frac{s^2}{\Lambda^4}-{\frac{c_{_6}}{2}}\frac{s^4}{\Lambda^8}$$
$$\therefore \mathcal{A}^-_{\text{tree}}=4c_{_2}\frac{s^2}{\Lambda^4}+{\frac{c_{_7}-c_{_6}}{2}}\frac{s^4}{\Lambda^8}~~\&~~\mathcal{A}^+_{\text{tree}}=8(2c_{_1}+c_{_2})\frac{s^2}{\Lambda^4}+{\frac{c_{_7}-2c_{_6}}{2}}\frac{s^4}{\Lambda^8}$$
We write the dim-12 tree contribution in the crossing symmetric amplitude as
$$\mathcal{A}_{\text{tree}}^{\pm,\text{dim-}12}\equiv \frac{s^4}{\Lambda^8}\sum_{I=6,7,8}T^I_{\pm}c_I$$
with 
$$T_+^6={-1},~T_-^6={-1/2}$$
$$T_+^7=T_-^7={1/2} $$
$$T_+^8=T_-^8={0}$$
\subsubsection*{UV divergences}
The photon loop has UV divergences $\sim s^4/\Lambda^8$, coming from two insertions of dim-8 operators in the loop:
\begin{equation}
    \mathcal{A}^{++++,+-+-}_{\text{uv}}=\frac{1}{\epsilon}\frac{7s^4}{20\pi^2\Lambda^8}(48 c_{_1}^2+40 c_{_1} c_{_2}+11 c_{_2}^2)
\end{equation}    
\begin{equation}
    \mathcal{A}^{++--}_{\text{uv}}=\frac{1}{\epsilon}\frac{5s^4}{3\pi^2\Lambda^8}(16 c_{_1}^2+16 c_{_1} c_{_2}+3 c_{_2}^2)
\end{equation}    
and the crossing symmetrized UV divergences are
\begin{equation}\label{Up}
    \mathcal{A}^{+}_{\text{uv}}=\frac{1}{\epsilon}\frac{ s^4}{60\pi^2\Lambda^8}\left(1808 c_{_1}^2+1640 c_{_1} c_{_2}+3
    81 c_{_2}^2\right) \equiv \frac{\mathcal{U}_+}{\epsilon}
\end{equation}
\begin{equation}\label{Um}
    \mathcal{A}^{-}_{\text{uv}}=\frac{1}{\epsilon}\frac{ s^4}{60\pi^2\Lambda^8}\left(208 c_{_1}^2+40 c_{_1} c_{_2}+81 c_{_2}^2\right)\equiv \frac{\mathcal{U}_-}{\epsilon}
\end{equation}
They need to be removed via dim-12 counter-terms $\delta c_I/\Lambda^8;~I=6,7,8$. Then the dim-12 contribution at tree level is
$$\mathcal{A}_{\text{tree}}^{\text{dim-}12}=\sum_{I}T_\pm^{I}c_I\frac{s^4}{\Lambda^8}+\sum_{I}T_\pm^{I}\delta c_I\frac{s^4}{\Lambda^8}+\mathcal{O}(t)$$
with the MS-scheme counter terms chosen as follows:
\begin{equation}\label{Upm}
    \sum_IT^I_\pm \delta c_I|_{MS}=-\mathcal{U}_\pm
\end{equation}
We will see that we only require this combination of counter-terms to compute the required RG running. Individual counter-terms are obtained as follows:
$$\frac{2\delta c_{_7}-3\delta c_{_6}}{4}\Big|_{MS}\frac{s^4}{\Lambda^8}=-\mathcal{A}^{++++}_{\text{uv}}, ~-\frac{\delta c_{_6}}{2}\Big|_{MS}\frac{s^4}{\Lambda^8}=-\mathcal{A}^{++--}_{\text{uv}},~\delta c_{_8}|_{MS}=0$$
$$\implies \delta c_{_6}|_{MS}=\frac{1}{\epsilon}\frac{10}{3\pi^2}(16c_{_1}^2+16c_{_1}c_{_2}+3c_{_2}^2),~~\delta c_{_7}|_{MS}=-\frac{1}{\epsilon}\frac{1}{5\pi^2}(48c_{_1}^2+20c_{_1}c_{_2}+16c_{_2}^2)$$$$\delta c_{_8}|_{MS}=0$$
In a general Renormalization Scheme,
$$\delta c_I = \delta c_I|_{MS} + f_I$$
We write this in a compact notation
$$\delta c_I \equiv -\frac{\mathcal{U}^I_{ij}c_ic_j}{\epsilon}+f_I$$
These are only computed discarding $\mathcal{O}(t)$ terms. \cite{BPS} compute them in more detail in Appendix D.

\subsubsection*{1-loop Finite Part}

\begin{equation}
    \mathcal{A}^{++++}_{\text{finite}}=\frac{7s^4}{40\pi^2\Lambda^8}\left((48c_{_1}^2+40c_{_1}c_{_2}+11c_{_2}^2)\log(\mu^2/s)\right) + \cdots
\end{equation}
\begin{equation}
    \mathcal{A}^{+-+-}_{\text{finite}}=\frac{7s^4}{40\pi^2\Lambda^8}\left((48c_{_1}^2+40c_{_1}c_{_2}+11c_{_2}^2)\log(\mu^2/s)\right) + \cdots
\end{equation}
\begin{equation}
    \mathcal{A}^{++--}_{\text{finite}}=\frac{5s^4}{6\pi^2\Lambda^8} \left((16 c_{_1}^2 + 16 c_{_1} c_{_2} + 3  c_{_2}^2)\log(\mu^2/s)\right)+ \cdots
\end{equation}
where $\cdots$ are non-$\log(s)$ terms ($\sim c_ic_js^4/\Lambda^8$), which are mentioned in Appendix \ref{EHres} for completeness. The finite part of the 1-loop crossing symmetric amplitude 
$$\mathcal{A}^-=\frac{1}{2}\left(\mathcal{A}^{++++}+\mathcal{A}^{+-+-}-\mathcal{A}^{++--}\right)$$
is
\begin{eqnarray*}
    \mathcal{A}^-_{\text{finite}} & =& -\frac{s^4}{240\pi^2\Lambda^8}\Big[\left(208 c_{_1}^2+40 c_{_1} c_{_2}+81 c_{_2}^2\right)\left( 2\log(s/\mu^2)-i\pi\right) - (15776/15)c_{_1}^2\\
    & & ~~~~~~~~~~~~~~~~- (560/15)c_{_1}c_{_2}
    - (4602/15)c_{_2}^2 \Big]
\end{eqnarray*}
$$\equiv \frac{s^4}{\Lambda^8}\left(A(2\log (s/\mu^2)-i\pi)+B\right)$$
For convenience, we have introduced the notation
\begin{equation}\label{ABdef}
    A \equiv -\frac{208c_{_1}^2+40c_{_1}c_{_2}+81c_{_2}^2}{240\pi^2},
    ~~B \equiv \frac{7888c_{_1}^2+280c_{_1}c_{_2}+2301c_{_2}^2}{1800\pi^2}
\end{equation}
So, after adding the dim-12 counter terms, we get the full forward amplitude (up to order $s^4/\Lambda^8$)
\begin{equation}\label{fullamp}
\mathcal{A}^-=4c_{_2}\frac{s^2}{\Lambda^4}+\sum_{I=6,7,8}T_-^{I}(c_I+f_I)\frac{s^4}{\Lambda^8}+\left[A(2\log (s/\mu^2)-i\pi)+B\right]\frac{s^4}{\Lambda^8}
\end{equation}
This is renormalization scheme dependent via $f_{_6},f_{_7},f_{_8}$ or simply $f_-\equiv \sum_{I} T_-^I f_I$.
Using
$$\int_{\cap_{s_0}}\frac{ds'}{i\pi}\frac{1}{s'}(A(2\log( s'/\mu^2)-i\pi) + B)= 2A\log (s_0/\mu^2) + B $$
the arc integral is
$$\int_{\cap_{s_0}}\frac{ds'}{i\pi}\frac{\mathcal{A}^-(s')}{s'^5}=0+\sum_{I=6,7,8}\frac{T_-^{I}c_I}{\Lambda^8}+\frac{f_-}{\Lambda^8}+\left[2A\log (s_0/\mu^2)+B\right]\frac{1}{\Lambda^8} $$
In Eq (D.13) of \cite{BPS} (arxiv version [v1]), they choose the Renormalization Scheme such that $f_{-}=-B$ so that the $s^4$ contribution comes only from dim-12 coefficients $c_I$. With this scheme, we reproduce their bound Eq (2.14):
$$\sum_IT_-^Ic_I+2A\log(s_0/\mu^2)>0$$
with $A$ defined in Eq (\ref{ABdef}).
We now show that the $\log s_0$ dependence cancels, generalizing the procedure explained in Sec \ref{simplercase}.\\ \\
Since
$$Im(\mathcal{A}^-)=+\frac{s^4}{240\pi\Lambda^8} \left(208 c_{_1}^2+40 c_{_1} c_{_2}+81 c_{_2}^2\right)=-\pi A \frac{s^4}{\Lambda^8}$$
we expect from Optical Theorem that the cross section will be (at order $s^3/\Lambda^8$)
$$\sigma(s)=\frac{s^3}{240\pi\Lambda^8} \left(208 c_{_1}^2+40 c_{_1} c_{_2}+81 c_{_2}^2\right)=-\pi A \frac{s^3}{\Lambda^8}$$
Using this, we can include the IR contribution from the RHS of the dispersion relation
$$\int_{\cap_{s_0}}\frac{ds'}{i\pi}\frac{\mathcal{A}^-(s')}{s'^5}=\frac{2}{\pi}\int_{s_0}^\infty ds' \frac{\sigma(s')}{s'^4}=\frac{2}{\pi}\int_{s_0}^{\Lambda^2} ds' \frac{\sigma(s')}{s'^4}+\frac{2}{\pi}\int_{\Lambda^2}^\infty ds' \frac{\sigma(s')}{s'^4}$$
The first term on the RHS can be evaluated within the EFT
$$\sum_I T_-^Ic_I\frac{1}{\Lambda^8}+2\frac{A}{\Lambda^8}\log(s_0/\mu^2)+\frac{B}{\Lambda^8}+\mathcal{O}(s_0/\Lambda^{10})=\frac{2}{\pi}\left(-\pi \frac{A}{\Lambda^8}\int_{s_0}^{\Lambda^2} ds'\frac{1}{s'}\right)+ \text{positive}$$
which cancels the $\log(s_0)$ piece on the LHS. Then we get
\begin{equation}
\sum_I T_-^Ic_I - 2|A|\log(\Lambda^2/\mu^2) > 0
\end{equation}
This bound is slightly better, since it is $s_0$-independent and the $\mathcal{O}(s_0/\Lambda^2)$ corrections can be dropped by taking $s_0\rightarrow 0$. The corrections introduced on the RHS are all positive due to Optical Theorem. Next, we compute the RG running of $c_I$'s and write the bound in terms of Wilson coefficients defined at the UV scale $\Lambda$. \\\\
We write the Lagrangian in terms of the renormalized fields and couplings:
$$A_\mu \rightarrow \sqrt{Z_A}A_\mu, ~c_i \rightarrow Z_i c_i, ~c_I \rightarrow Z_I c_I$$
where $i=1,2$ labels dim-8 operators and $I=6,7,8$ labels dim-12 operators.
$$Z_i=1+\delta_i,~Z_I=1+\delta_I$$
We will have combinations of the type $$Z_A^2Z_i=(1+\delta_A)^2(1+\delta_i)= 1 + (2\delta_A+\delta_i)+\mathcal{O}(\delta^2)$$
where we can neglect $\delta_A^2$ terms as they are $\mathcal{O}(p^8/\Lambda^8)$ and contribute to $\delta c/\Lambda^4$ at $\mathcal{O}(p^{12}/\Lambda^{12})$. Note that $\delta_A\sim p^4/\Lambda^4$ contributes to $\delta c/\Lambda^4$ at $\mathcal{O}(p^8/\Lambda^8)$. So, we only need $\delta_A$ to $\mathcal{O}(p^4/\Lambda^4)$, which comes from a scaleless loop and hence vanishes in dim-reg. So we can take $\delta_A=0$ (see also footnote \ref{WFR}).
\begin{eqnarray*}
    \mathcal{L} & = & -\frac{1}{4}F_{\mu\nu}F^{\mu\nu}+\frac{c_{_1}}{\Lambda^4}(F_{\mu\nu}F^{\mu\nu})^2+\frac{c_{_2}}{\Lambda^4}F_{\mu\nu}F^{\nu\sigma}F_{\sigma\rho}F^{\rho\mu} \\
    & & +\frac{c_{_6}}{\Lambda^8}\partial_\nu F_{\alpha \rho} \partial^\alpha F^{\nu \sigma} \partial^\delta F^{\beta \rho} \partial_\beta F_{\delta \sigma}+
    \frac{c_{_7}}{\Lambda^8} F_{\rho \beta} \partial_\alpha F_{\sigma \gamma} \partial^{\beta} F^{\sigma \delta} \partial_{\delta} \partial^{\gamma} F^{\rho \alpha}
    +\frac{c_{_8}}{\Lambda^8} F_{\alpha \beta} F_{\sigma \gamma} \partial^\alpha \partial_{\rho} F^{\gamma \delta} \partial^\beta \partial_{\delta} F^{\rho \sigma} \\
    & & -\delta_A\frac{1}{4}F_{\mu\nu}F^{\mu\nu}+(2\delta_A+\delta_1)\frac{c_{_1}}{\Lambda^4}(F_{\mu\nu}F^{\mu\nu})^2+(2\delta_A+\delta_2)\frac{c_{_2}}{\Lambda^4}F_{\mu\nu}F^{\nu\sigma}F_{\sigma\rho}F^{\rho\mu} \\
    && +(2\delta_A+\delta_6)\frac{c_{_6}}{\Lambda^8}\partial_\nu F_{\alpha \rho} \partial^\alpha F^{\nu \sigma} \partial^\delta F^{\beta \rho} \partial_\beta F_{\delta \sigma}+
    (2\delta_A+\delta_7)\frac{c_{_7}}{\Lambda^8} F_{\rho \beta} \partial_\alpha F_{\sigma \gamma} \partial^{\beta} F^{\sigma \delta} \partial_{\delta} \partial^{\gamma} F^{\rho \alpha}\\
    && +(2\delta_A+\delta_8)\frac{c_{_8}}{\Lambda^8} F_{\alpha \beta} F_{\sigma \gamma} \partial^\alpha \partial_{\rho} F^{\gamma \delta} \partial^\beta \partial_{\delta} F^{\rho \sigma} 
    \end{eqnarray*}
So, 
\begin{equation}\label{ctdef}
    \delta c_I \equiv (2\delta_A + \delta_I)c_I,~~\delta c_i \equiv (2\delta_A + \delta_i)c_i
\end{equation} 
Since $\delta_A=0$, 
$$\delta c_I=\delta_Ic_I,~~\delta c_i=\delta_ic_i$$
RG running of $c_I$'s is given by the RGEs: $$\mu\frac{d}{d\mu}(\mu^\epsilon Z_I c_I)=\mu\frac{d}{d\mu}(\mu^\epsilon (c_I + \delta c_I))=0$$ 
$$\implies \epsilon (c_I+\delta c_I)+\mu\frac{d}{d\mu}(c_I+\delta c_I)=0$$
There can't be a dim-8 counter term $\frac{\delta c_i}{\Lambda^4}F^4$ since this would require a loop with two insertions of dim-6 operators $\frac{c}{\Lambda^2}\mathcal{O}_{\text{dim-6}}$ which don't exist for this theory. So the $\beta$ function of the dim-8 coefficients $c_i$ is obtained by 
$$\mu\frac{d}{d\mu}(\mu^\epsilon c_i)=0$$
\begin{equation}\label{betaci}
\implies \mu\frac{d}{d\mu}c_i=-\epsilon c_i~~~~~
\end{equation}
i.e. $c_i$'s don't run.
So, 
\begin{eqnarray}
    \mu\frac{d}{d\mu}\delta c_I&=&\mu\frac{d}{d\mu}\left(-\frac{\mathcal{U}^I_{ij}c_ic_j}{\epsilon}+f_I\right) \nonumber\\
    &=&-2\frac{\mathcal{U}^I_{ij}c_i}{\epsilon}\mu\frac{d}{d\mu}c_j+\mu\frac{d}{d\mu}f_I \label{int}\\
    &=&+2\mathcal{U}^I_{ij}c_ic_j+\mu\frac{d}{d\mu}f_I \nonumber
\end{eqnarray}
Note that in Eq (\ref{int}) we need the $\beta$ function of $c_i$ to $\mathcal{O}(\epsilon)$ (Eq (\ref{betaci})).
$$\therefore \epsilon (c_I+\delta c_I)+\mu\frac{d}{d\mu}(c_I+f_I)+2\mathcal{U}^I_{ij}c_ic_j=0$$
Setting terms of $\mathcal{O}(\epsilon)=0$, we get
$$\mu\frac{d}{d\mu}(c_I+f_I)(\mu)=(-2+1)\mathcal{U}^I_{ij}c_ic_j|_{\mu_0}$$
and $$\mu\frac{d}{d\mu}\sum_I T_-^I (c_I+f_I)(\mu)=-\sum_I\mathcal{U}^I_{ij}c_ic_j=-\mathcal{U}_-~~~\text{(from Eq (\ref{Um}) and (\ref{Upm}))}$$
Integrating for $C\equiv \sum_I T_-^I(c_I+f_I)$, we get
$$C(\mu)-C(\mu_1)=-\mathcal{U}_-\log(\mu/\mu_1)$$
or
$$C(\mu)=C(\Lambda)+\mathcal{U}_-\log(\Lambda/\mu)$$
where we have chosen $\mu_1=\Lambda$, where the EFT is matched to the UV theory.
Then, the bound on $c$ can be written as
$$c(\Lambda)+ f_-(\Lambda)+\mathcal{U}_-\log(\Lambda/\mu)-4|A|\log(\Lambda/\mu) + B +\mathcal{O}(\epsilon) >0$$
$$\mathcal{U}_-=\frac{1}{60\pi^2}\left(208c_{_1}^2+40c_{_1}c_{_2}+81c_{_2}^2\right)$$
$$4|A|=\frac{208c_{_1}^2+40c_{_1}c_{_2}+81c_{_2}^2}{60\pi^2}$$
With the scheme $f_-=-B$, we have the bound at $\Lambda$
\begin{equation}\label{finalbound1}
\sum_IT_-^Ic_I(\Lambda)>0
\end{equation}
We can also substitute the running behaviour in Eq (\ref{fullamp}) to see that the amplitude is indeed $\mu$-independent.
$$\mathcal{A}^-(s)=4c_{_2}\frac{s^2}{\Lambda^4}+\sum_{I=6,7,8}T_-^{I}(c_I+f_I)(\Lambda)\frac{s^4}{\Lambda^8}+\left[A(2\log (s/\Lambda^2)-i\pi)+B\right]\frac{s^4}{\Lambda^8}$$
One can do the same analysis with this amplitude and arrive at the final bound Eq (\ref{finalbound1}).\\ \\
The same analysis can be repeated for the crossing symmetric amplitude with the different sign
$$\mathcal{A}^+=\frac{1}{2}\left(\mathcal{A}^{++++}+\mathcal{A}^{+-+-}\textcolor{red}{+}\mathcal{A}^{++--}\right)$$
This amplitude is
$$\mathcal{A}^+=8(2c_{_1}+c_{_2})\frac{s^2}{\Lambda^4}+\sum_{I=6,7,8}T_+^{I}(c_I+f_I)(\Lambda)\frac{s^4}{\Lambda^8}+\left[A'(2\log (s/\Lambda^2)-i\pi)+B'\right]\frac{s^4}{\Lambda^8}$$
where
\begin{eqnarray}
    A'&\equiv& -\frac{1808c_{_1}^2+1640c_{_1}c_{_2}+381c_{_2}^2}{240\pi^2}\nonumber\\
    B'&\equiv&\frac{21424c_{_1}^2+22360c_{_1}c_{_2}+5583c_{_2}^2}{1800\pi^2}\label{A'B'def}
\end{eqnarray}
and the final bound is
\begin{equation}\label{finalbound2}
\sum_IT_+^Ic_I(\Lambda)>0
\end{equation}
in the scheme $\sum_IT^I_+f_I=-B'$. So, we again see that the only modifications at 1-loop are due to RG effects.
\section{EFT of Gluons}\label{QCDEH}
In this section, we consider EFT of Gluons, in particular dim-6 and dim-8 operators. The complete list of dim-6 operators is given in \cite{dim6} and that for dim-8 operators in \cite{dim81, dim82}, which we present in Table \ref{dim6_dim8_operators}.
$$\mathcal{L}=-\frac{1}{4}G^a_{\mu\nu}G^{a\mu\nu}+\sum_{i=1,2}\frac{c_{_6}^{(i)}}{\Lambda^2}G^3_{(i)}+\sum_{i=1,..,9}\frac{c_{_8}^{(i)}}{\Lambda^4}G^4_{(i)}$$
For simplicity, we consider the theory with only $c_{_8}^{(1)}, c_{_8}^{(3)}\neq 0$.
\begin{table}[h]
\centering
\resizebox{12cm}{!}{
\begin{tabular}{||c|c||c|c||}
\hline \multicolumn{2}{||c||}{$G^{3}$} & \multicolumn{2}{c||}{$G^{4}$} \\
\hline
$G^3_{(1)}$ &$f^{abc} G_\mu^{a\nu}G_\nu^{b\rho}G_\rho^{c\mu}$&$G^4_{(1)}$ & $\left(G_{\mu \nu}^{a} G^{a \mu \nu}\right)\left(G_{\rho \sigma}^{b} G^{b \rho \sigma}\right)$   \\
$G^3_{(2)}$ & $f^{abc} \widetilde{G}_\mu^{a\nu}G_\nu^{b\rho}G_\rho^{c\mu}$&$G^4_{(2)}$ & $\left(G_{\mu \nu}^{a} \widetilde{G}^{a \mu \nu}\right)\left(G_{\rho \sigma}^{b} \widetilde{G}^{b \rho \sigma}\right)$  \\
 &  & $G^4_{(3)}$ &  $\left(G_{\mu \nu}^{a} G^{b \mu \nu}\right)\left(G_{\rho \sigma}^{a} G^{b \rho \sigma}\right)$\\
 &  & $G^4_{(4)}$ & $\left(G_{\mu \nu}^{a} \widetilde{G}^{b \mu \nu}\right)\left(G_{\rho \sigma}^{a} \widetilde{G}^{b \rho \sigma}\right)$ \\
 &  &$G^4_{(5)}$ & $\left(G_{\mu \nu}^{a} G^{a \mu \nu}\right)\left(G_{\rho \sigma}^{b} \widetilde{G}^{b \rho \sigma}\right)$ \\
&  & $G^4_{(6)}$ & $\left(G_{\mu \nu}^{a} G^{b \mu \nu}\right)\left(G_{\rho \sigma}^{a} \widetilde{G}^{b \rho \sigma}\right)$ \\
 &  & $G^4_{(7)}$ & $d^{a b c} d^{d e c}\left(G_{\mu \nu}^{a} G^{b \mu \nu}\right)\left(G_{\rho \sigma}^{d} G^{e \rho \sigma}\right)$\\
 &  & $G^4_{(8)}$ & $d^{a b c} d^{d e c}\left(G_{\mu \nu}^{a} \widetilde{G}^{b \mu \nu}\right)\left(G_{\rho \sigma}^{d} \widetilde{G}^{e \rho \sigma}\right)$\\
 &  & $G^4_{(9)}$ & $d^{a b c} d^{d e c}\left(G_{\mu \nu}^{a} G^{b \mu \nu}\right)\left(G_{\rho \sigma}^{d} \widetilde{G}^{e \rho \sigma}\right)$\\
\hline
\end{tabular}
}
 \caption{Dimension 6 and 8 gluonic operators}
 \label{dim6_dim8_operators}
\end{table}
The tree level bounds on this theory were derived in \cite{Ghosh:2022qqq}. We wish to see the effect on the bounds due to 1-loop QCD corrections i.e. at $\mathcal{O}(g^4)$. Dimension-8 EFT coefficients can only obtained from a loop level UV completion i.e. $c_{_8}\sim g^4/(16\pi^2)$ 
 (see \cite{Quevillon:2018mfl}) 
is generated at 1-loop.
In this case, the QCD-loop and dim-8 tree are at the same order in $g$. So the 1-loop analysis is necessary. We consider  $2\rightarrow 2$ scattering
of gluons, all with the same colour state $(a=b=c=d)$. This is also convenient as it gets rid of the UV divergence in the Amplitude. But unlike the photon case, the 1-loop amplitude contains IR and $\log t$ divergences. We regulate both of them by assigning a mass $M$ to gluons and taking the limit $M\rightarrow 0$ at the end.
\begin{figure}[h!]
    \centering
    \includegraphics[width=0.5\linewidth]{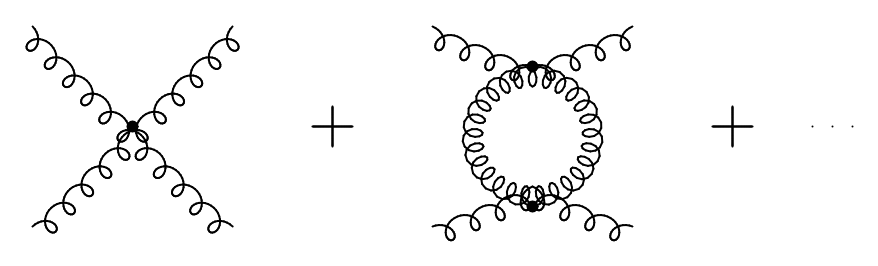}
    \caption{Left: dim-8 contact interaction, Right: QCD interactions}
    \label{fig:gluons} 
\end{figure}\\
The forward scattering amplitude gets contribution from dim-8 contact interactions at tree level (first term) and from QCD loop (second term) at order $g^4$ (see Fig \ref{fig:gluons}):\footnote{$C_0(s)$ is short hand for the Passarino-Veltman function $C_0(M^2,M^2,s,M^2,M^2,M^2)$, which has a complicated expression in terms of polylogarithms, and is defined as follows:
$$C_0(k_1^2,k_2^2,(k_1+k_2)^2,m_1^2,m_2^2,m_3^2)=\int\frac{d^Dq}{(2\pi)^D}\frac{1}{(q^2-m_1^2)\left((q+k_1)^2-m_2^2\right)\left((q+k_1+k_2)^2-m_3^2\right)}$$}
\begin{eqnarray}
    \mathcal{A}_{\lambda_1\lambda_2\lambda_3\lambda_4}(s) & = & 2\big(c_{_8}^{(1)}+c_{_8}^{(3)}\big)\frac{s^2}{\Lambda^4}\Big(\lambda_1(2\lambda_2\lambda_3\lambda_4+\lambda_2-\lambda_4)-\lambda_2\lambda_3+\lambda_3\lambda_4+2\Big) \nonumber\\ 
    & &+ \frac{g^4f^4}{16\pi^2}\Big[(\lambda_1\lambda_3+1)(\lambda_2\lambda_4+1)\left(i\pi\frac{s}{M^2}-\log(s/M^2)\right)\nonumber\\ \nonumber\\
    & &-2(\lambda_1+\lambda_3)(\lambda_2+\lambda_4)s\left(C_0(s)+C_0(4M^2-s)\right)\Big]  + \text{finite}
\end{eqnarray} 
where
\begin{eqnarray}
    \text{finite} & = & -\frac{g^4f^4}{96\pi^2}i\Big[\lambda_1(\lambda_2(i(-9 + 3i\pi + 14\sqrt3\pi)\lambda_3\lambda_4 + 
                27\pi - 3i) + i(-3 + 3i\pi + 
                14\sqrt3\pi)\lambda_3 + 3(9\pi + i)\lambda_4)\nonumber \\
    & & +3(9\pi + i)\lambda_2\lambda_3 + 
      i(-3 + (14\sqrt3 + 3i)\pi)\lambda_2\lambda_4 - 
      3i\lambda_3\lambda_4 + 27\pi\lambda_3\lambda_4 + 14i\sqrt3\pi - 
      3\pi - 9i \Big] \nonumber
\end{eqnarray}
and  $f^4\equiv f^{aij}f^{a}_{~ik}f^{a}_{~jl}f^{akl}>0$. This is the complete expression in the $M\rightarrow 0$ limit.\\ \\
The crossing-symmetric amplitude\footnote{{As mentioned in Sec \ref{gen}, this was shown for photons in \cite{BPS}. But since we have taken all colour states to be the same, we assume this for gluons as well.}}
$$\mathcal{A}(s)=\frac{\mathcal{A}_{++++}+\mathcal{A}_{+-+-}-\mathcal{A}_{++--}}{2}$$
is
\begin{equation}
    \mathcal{A}(s)=8\big(c_{_8}^{(1)}+c_{_8}^{(3)}\big)\frac{s^2}{\Lambda^4}+\frac{g^4f^4}{16\pi^2}\left[4\left(i\pi\frac{s}{M^2}-\log(s/M^2)\right)+2i\pi+28\pi/\sqrt{3}-2\right]
\end{equation}
The usual positivity analysis 
$$\int_{\cap_{s_0}}\frac{ds'}{i\pi}\frac{\mathcal{A}(s')}{s'^3}=\frac{2}{\pi}\int_{s_0}^\infty ds' \frac{\sigma(s')}{s'^2}>0$$
leads to
\begin{equation}
    \frac{8}{\Lambda^4}\big(c_{_8}^{(1)}+c_{_8}^{(3)}\big)+\frac{g^4f^4}{16\pi^2}\left[4\left(\frac{2}{s_0M^2}+\frac{1}{2s_0^2}\right)+0\right]>0
\end{equation}
which would lead us to believe that the tree level bound ($c_{_8}^{(1)}+c_{_8}^{(3)}>0$ as obtained in \cite{Ghosh:2022qqq}) has a ``divergent correction" (when $M\rightarrow 0$). \\ \\
We notice that the cross-section is a sum of cross-sections over final states and break it as
\begin{eqnarray*}
    \sigma(s)&=&\sigma(gg\rightarrow gg)+\sum_{X\neq gg}\sigma(gg\rightarrow X) 
\end{eqnarray*}
Since the second term receives contribution from $s>\Lambda^2$, the first term can be obtained by Optical Theorem for $s<\Lambda^2$, at order $g^4$: 
\begin{equation}\label{g4cs}
    Im\mathcal{A}(s)=s\sigma(s)=\frac{g^4f^4}{16\pi^2}\left[4\pi\frac{s}{M^2}+2\pi\right]+ ...>0~ ~ (s<\Lambda^2)
\end{equation}
where dots indicate contribution from processes at higher order in $g$ and/or $1/\Lambda$. Since dim-8 coefficients are generated at least at 1-loop, this expression is expected to hold for $s>\Lambda^2$ as well i.e. the cross section for $gg\rightarrow gg$ at $g^4$ order is 
\begin{equation}\label{g4_cs_all_s}
    \sigma(gg\rightarrow gg)|_{g^4}=\frac{g^4f^4}{8\pi}\left[\frac{2}{M^2}+\frac{1}{s}\right] ~~~\forall s 
\end{equation}
So, we include this contribution from the RHS:
\begin{equation*}
    \int_{\cap_{s_0}}\frac{ds'}{i\pi}~\frac{\mathcal{A}(s')}{s'^{3}}-\frac{2}{\pi}\int_{s_0}^{\infty} ds' \frac{\sigma(gg\rightarrow gg)|_{g^4}+\mathcal{O}(g^6)}{s'^{2}} = \frac{2}{\pi}\int_{s_0}^{\infty} ds' \frac{\sum_{X\neq gg}\sigma(gg\rightarrow X)}{s'^{2}}>0
\end{equation*}
$$ \implies \frac{8}{\Lambda^4}\big(c_{_8}^{(1)}+c_{_8}^{(3)}\big)+\frac{g^4f^4}{4\pi^2}\left(\cancel{\frac{2}{s_0M^2}+\frac{1}{2s_0^2}}\right)-\frac{2}{\pi}\frac{g^4f^4}{8\pi}\left[\left(\cancel{\frac{2}{s_0M^2}+\frac{1}{2s_0^2}}\right)-0\right]>0$$
where the `$0$' in the last bracket comes due to integrating up to infinity\footnote{If we only integrate up to $\Lambda$, we are still left with a $+1/\Lambda^2M^2$ divergence on the LHS.}. So,  we recover the tree level bound, with corrections of $\mathcal{O}(g^6)$:
\begin{equation}
    c_8^{(1)}+c_8^{(3)}+\mathcal{O}(g^6,s_0/\Lambda^2)>0 
\end{equation}
The $s_0/\Lambda^2$ suppression can be minimized by taking $s_0<<\Lambda^2$ \footnote{Since asymptotic gluon states are defined at energies $>\Lambda_{QCD}$, we take $$\Lambda_{QCD}^2<<s_0<<\Lambda^2$$}, but one has to do a higher loop analysis to say something at higher orders in $g$.

\section{Summary}\label{ext}

Positivity bounds impose constraints on the coefficients of EFT operators,
based on fundamental properties of the underlying UV-complete theory.
While tree-level bounds are well-established and derived in various contexts,
the loop contribution can often occur at the same order as the tree level
contribution from higher dimensional operators. Consequently, it is
essential to extend the understanding of positivity bounds to loop-level contributions. Although loop-level analyses have been explored in the literature, they often encounter challenges such as infrared (IR) logs, IR divergences, and other complications. In this paper, we address some of these challenges for specific theories.\\\\
In Sec \ref{EH12}, we showed how one can include IR scale ($s_0$)-independent loop corrections to positivity bounds (derived from forward dispersion relations) in the EFT of Photons, and get an exact bound on the dim-12 Wilson coefficients. We concluded that there are no ``non-trivial" loop corrections to the tree level bound, the trivial ones coming from RG effects. In other words, the tree level bound $$\sum_IT^Ic_I>0$$ after including loop corrections (in our case, just one loop sufficed) becomes $$\sum_IT^Ic_I(\mu)+2A\log(\Lambda^2/\mu^2)+f>0$$ where the log term  comes from the RG evolution of $c_I$'s, and $f$, an arbitrary finite piece, comes from Renormalization Scheme dependence, which can be fixed once the couplings are defined. At the EFT matching scale, we simply get 
$$\sum_IT^Ic_I(\Lambda)>0$$
i.e. the tree level bound continues to hold, even when the loop contribution is taken into account.\\ \\
The procedure followed in Sec \ref{EH12} for the Photon EFT case may be generalized to derive bounds on higher dimensional coefficients and to other EFTs with well defined forward limit and IR safe loop amplitudes, if the only perturbative parameter is $\partial/\Lambda$ (unlike the gluon case), which we outline below:
\begin{itemize}
    \item[1.] Write the dispersion relation by completely separating UV and IR contributions, with a crossing symmetric amplitude
    $$\int_{\cap_{s_0}}\frac{ds}{i\pi}~\frac{\mathcal{A}_{\text{EFT}}(s)}{s^{2n+1}}=\frac{2}{\pi}\int_{s_0}^{\Lambda^2} ds \frac{\sigma(s)}{s^{2n}}+\frac{2}{\pi}\int_{\Lambda^2}^{\infty} ds \frac{\sigma(s)}{s^{2n}}$$
    \item[2.] To derive an exact bound on the dimension-$\mathcal{D}$ Wilson coefficient, separate the amplitude based on powers of $(s/\Lambda^2)$ as follows: $$\mathcal{A}_{\text{EFT}}= \mathcal{A}_{< 2n}+\mathcal{A}_{2n}+\mathcal{A}_{>2n}$$
    where the subscripts denote the powers of $s/\Lambda^2$ and $\mathcal{D}=4(n+1)$\footnote{So, one can use this analysis only to bound coefficients with $\mathcal{D}=4(n+1)$ i.e. multiples of 4. This is because power of $s$ must be even for $s$-$u$ symmetry.}. Explicitly,
    $$\mathcal{A}_{2n}=\left[(\text{dim}<\mathcal{D})+c_{\mathcal{D}}\right]\left(\frac{s}{\Lambda^2}\right)^{2n}$$
    $(\text{dim}<\mathcal{D})$ denotes the loop contributions from lower dimensional operators that contribute at the same order as the dim-$\mathcal{D}$ tree.
    \item[3.] Choosing $n=(\mathcal{D}-4)/4$ to pick out the coefficient $c_{\mathcal{D}}$, and further separating the IR cross-section based on powers of $s/\Lambda^2$, the dispersion relation becomes
    \begin{eqnarray*}
        \int_{\cap_{\textcolor{red}{{s_0}}}}\frac{ds}{i\pi}\frac{\mathcal{A}_{<2n}}{s^{2n+1}} +\frac{c_{\mathcal{D}}}{\Lambda^{\mathcal{D}-4}}+\mathcal{O}\left(\textcolor{blue}{s_0}\over \Lambda^{\mathcal{D}-2}\right)-\frac{2}{\pi}\int_{\textcolor{red}{s_0}}^{\Lambda^2} ds \frac{Im\mathcal{A}_{\leq 2n}}{s^{2n+1}}&=&\frac{2}{\pi}\int_{\textcolor{blue}{s_0}}^{\Lambda^2} ds \frac{Im\mathcal{A}_{>n}}{s^{2n+1}}\\
        &&+\frac{2}{\pi}\int_{\Lambda^2}^{\infty} ds \frac{\sigma(s)}{s^{2n}}
    \end{eqnarray*}
    Taking $s_0\rightarrow 0$ minimizes the third term (corrections from higher dimensional operators) and may give rise to divergences in the first term, generically of the type
    $$\frac{1}{s_0^{2n}}\left(\frac{s_0}{\Lambda^2}\right)^{2n-2k}\times\log^p(s_0)=\frac{1}{s_0^{2k}\Lambda^{4(n-k)}}\times \log^p(s_0) $$ But, since $s_0$ dependence must cancel \textit{order by order} in $1
    /\Lambda$,
    it should cancel between first term and fourth term from LHS (in red). So, there will be no divergences due to $s_0\rightarrow 0$. Of course, the remaining $s_0$ dependence (in blue) from third term on LHS and first term on RHS  (higher order in $p/\Lambda$) should also cancel separately (which we cannot check anyway, since we don't know contributions from higher dimensional operators) but we keep it split this way so that the corrections (on LHS) can be minimized by $s_0\rightarrow 0$ and utilize the fact that RHS is positive. 
    \item[4.] So, after first cancelling $s_0$ dependence from lower dimensional contributions and then taking $s_0\rightarrow 0$, we get a bound 
    $$\lim_{s_0\rightarrow 0^+} \left[c_{\mathcal{D}}+\Lambda^{\mathcal{D}-4}\left(\int_{\cap_{{{s_0}}}}\frac{ds}{i\pi}\frac{\mathcal{A}_{<2n}}{s^{2n+1}}-\frac{2}{\pi}\int_{s_0}^{\Lambda^2} ds \frac{Im\mathcal{A}_{\leq 2n}}{s^{2n+1}}\right)\right]>0$$ Of course, for bounds on higher dimension coefficients, one needs to evaluate higher number of loops to get the exact contribution $\mathcal{A}_{\leq 2n}$. $s_0\rightarrow 0$ gets rid of $\mathcal{O}(s_0/\Lambda^2)$ corrections. But if there are massive particles in the EFT ($m<<\Lambda$) or marginal couplings (say $g$), the bound can get $\mathcal{O}(m^2/\Lambda^2)$ or $\mathcal{O}(g^p)$ corrections respectively.\\
\end{itemize}
In Sec \ref{QCDEH}, we considered EFT of Gluons where the forward limit is not well defined at 1-loop and the procedure outlined above does not apply. However, we showed that the $\log t$/IR divergence could be cancelled by analysing the RHS, to obtain a bound that is divergence free, at least at the order $g^4$. \\ \\
Positivity bounds provide a non-trivial way to constrain the space of EFTs that are compatible with a Unitary, Lorentz Invariant and local/causal UV completion. So it is important to make sure that they are meaningful even at loop level. Interesting future directions can be to understand loop effects on positivity bounds derived from near forward/non-forward dispersion relations, making optimal use of the RHS in the EFT regime, while also minimizing the higher derivative corrections.
\appendix 
\section{EFT of Photons: dim-8 loop}\label{EHres}
Here, we present the loop computation results from Sec \ref{EH12}. These were computed both with and without mass-regulator; the $\log M^2$'s and $\log (-t)$'s cancel respectively at the end, in both cases, unlike the gluon case where we have IR and $\log t$ divergences.
\begin{eqnarray*}
    \mathcal{A}_{\text{loop}}^{++++}&=& \frac{7s^4}{40\pi^2\Lambda^8}\left(\left(48c_{_1}^2+40c_{_1}c_{_2}+11c_{_2}^2\right)\left(\frac{2}{\epsilon}+\log(\mu^2/s)\right)\right) \\ 
    && +\frac{s^4}{1200\pi^2\Lambda^8}\Big((16c_{_1}^2(570i\pi+1057)+120c_{_1}c_{_2}(58i\pi+125)\\
    &&~~~~~~~~~~~~~~~~~~~+c_{_2}^2(1770i\pi+4489))\Big)
\end{eqnarray*}
\begin{eqnarray*}
    \mathcal{A}_{\text{loop}}^{+-+-}&=&\frac{7s^4}{40\pi^2\Lambda^8}\left(\left(48c_{_1}^2+40c_{_1}c_{_2}+11c_{_2}^2\right)\left(\frac{2}{\epsilon}+\log(\mu^2/s)\right)\right) \\
    && +\frac{s^4}{1200\pi^2\Lambda^8}\Big((16c_{_1}^2(60i\pi+1057)+120c_{_1}c_{_2}(12i\pi+125) \\
    &&~~~~~~~~~~~~~~~~~~~+c_{_2}^2(540i\pi+4489))\Big)
\end{eqnarray*}
\begin{eqnarray*}
    \mathcal{A}_{\text{loop}}^{++--}&=&\frac{5s^4}{6\pi^2\Lambda^8} \left(\left(16 c_{_1}^2 + 16 c_{_1} c_{_2} + 3  c_{_2}^2\right)\left(\frac{2}{\epsilon}+\log(\mu^2/s)\right)\right) \\
    && +\frac{s^4}{360\pi^2\Lambda^8}\Big((16c_{_1}^2(150i\pi+437)+8c_{_1}c_{_2}(300i\pi+1111) \\
    &&~~~~~~~~~~~~~~~~~~ +9c_{_2}^2(50i\pi+197))\Big)
\end{eqnarray*}

\section{Comments on Eq (\ref{finalboundsimp})}\label{comment}
In this appendix, we give some comments on the bound derived in Sec \ref{simplercase}, Eq (\ref{finalboundsimp})$$c_{_7}(\Lambda)>0$$ This was derived by taking into account the integral on RHS of dispersion relation up to the EFT cut-off scale $\Lambda$. One could have done the same analysis by breaking the integral up to some lower scale $\Tilde{\Lambda}<\Lambda$
i.e.
$$\int_{\cap_{s_0}}\frac{ds}{i\pi}~\frac{\mathcal{A}(s)}{s^5}=\frac{2}{\pi}\int_{s_0}^{\Tilde{\Lambda}^2} ds \frac{\sigma(s)}{s^4}+\frac{2}{\pi}\int_{\Tilde{\Lambda}^2}^{\infty} ds \frac{\sigma(s)}{s^4}$$
Then, following the same steps as in Sec \ref{simplercase}, instead of Eq (\ref{simple bound}), we will get
\begin{equation}\label{simple bound Tilde}
    c_{_7}(\mu)+f_{_7}-\frac{27c_{_2}^2}{20\pi^2}\log(\Tilde{\Lambda}^2/\mu^2)>0
\end{equation} 
To be precise, we have
\begin{equation}\label{IPATildeLambda}
    c_{_7}(\mu)+f_{_7}-\frac{27c_{_2}^2}{20\pi^2}\log(\Tilde{\Lambda}^2/\mu^2)=\frac{2\Lambda^8}{\pi}\int_{\Tilde{\Lambda}^2}^{\Lambda^2} ds \frac{\sigma(s)}{s^4}+\frac{2\Lambda^8}{\pi}\int_{\Tilde{\Lambda}^2}^{\infty} ds \frac{\sigma(s)}{s^4} 
\end{equation} 
Choosing the scheme $f_{_7}=0$\footnote{This can be restored by $c\rightarrow c+f$. Any statement about $c$ will turn into a statement about $c+f$.}, and by the RG running of $c_{_7}$ (Eq (\ref{c7running})), we get
$$c_{_7}(\Tilde{\Lambda})=\frac{2\Lambda^8}{\pi}\int_{\Tilde{\Lambda}^2}^{\Lambda^2} ds \frac{\sigma(s)}{s^4}+\frac{2\Lambda^8}{\pi}\int_{\Tilde{\Lambda}^2}^{\infty} ds \frac{\sigma(s)}{s^4} >0$$
Since the choice of $\Tilde{\Lambda}$ was arbitrary and the scheme choice was also independent of $\Tilde{\Lambda}$, it seems that at any lower energy scale, the analysis gives $$c_{_7}(\Tilde{\Lambda})>0$$ 
On the other hand, this also follows from positivity of $c_{_7}(\Lambda)$ and RG running of $c_{_7}$:
$$c_{_7}(\mu)=c_{_7}(\Lambda)+\frac{27c_{_2}^2}{10\pi^2}\log(\Lambda/\mu)>c_{_7}(\Lambda)>0$$
From the RG perspective, this is true only because of the positive coefficient in front of $\log(\Lambda/\mu)$, while from the analysis perspective,
this is happening because the RG running log is provided by the first term of Eq (\ref{IPATildeLambda}), which is positive due  to Optical Theorem. But the strongest bound is obtained once we take the full EFT contribution from RHS, which in this case just gives $$c_{_7}(\Lambda)>0$$
One would think that if in some case it turned out that the coefficient was negative, positivity would not have been preserved at lower energy scales. But we argue below that its sign is fixed to be positive by Optical Theorem and crossing symmetry. \\ \\
Consider the same form of Amplitude that we had:
$$\mathcal{A}(s)=\left(\frac{s^2}{\Lambda^4}\right)^n\mu^\epsilon\left(c(\mu)+(2\log(s/\mu^2)-i\pi)F + \text{real}\right) $$
This is the most general crossing-symmetric Amplitude in absence of IR/$\log t$ divergences and additional mass scales or marginal couplings, at 1-loop (see footnote \ref{weak_coupling}). The log and $i\pi$ coefficients get related due to crossing symmetry of the Amplitude and Optical Theorem (at this order) fixes the sign of $F$:
$$Im\mathcal{A}=-\left(\frac{s^2}{\Lambda^4}\right)^n\pi F>0 \implies F<0$$
Since RG running is determined by $$\mu\frac{d}{d\mu}\mathcal{A}=0$$
it is not surprising that the sign of RG running will be determined by that of $F$, which is fixed by Optical Theorem. RG running of $c$ up to 1-loop can be computed\footnote{Note that for higher dimensional coefficients, $F$ (proportional to square of lower dimensional coefficients) will run in general. To ignore its running, we have assumed that all coefficients $c\sim g_{weak}$, so that $F\sim g_{weak}^2$ and $$F(\mu)=F(\Lambda)+\mathcal{O}(g_{weak}^3)$$ \label{weak_coupling}}:
$$\epsilon(...)+\mu^\epsilon\left[\mu\frac{dc(\mu)}{d\mu}-4F\right]=0\implies c(\mu)-c(\Lambda)=4F\log(\mu/\Lambda)$$
Hence, if $c(\Lambda)>0$, the sign is preserved:
$$c(\mu)=c(\Lambda)+ (-4F)\log(\Lambda/\mu)>0 + 0$$
Interestingly, the anomalous dimension $4F$ can be computed from the imaginary part of the Amplitude, which can be determined from products of tree level amplitudes using the usual Cutkosky cutting rules
$$F\sim Im\mathcal{A}\sim\sum_{\text{cuts}}(\text{tree})\times(\text{tree})$$
 See \cite{RG_from_tree_Caron-Huot:2016cwu,RG_from_tree_Baratella:2020lzz, RG_from_tree_EliasMiro:2020tdv} and the references therein, where this idea is explored more generally. For our example, the crossing symmetry was required to relate the imaginary part and $\log\mu$ coefficients\footnote{The well defined forward limit of the Amplitude was also important; this is not straightforwardly true in the QCD case, due to the forward limit being not well defined at 1-loop, introducing $\log(s/t)$ terms.}. 
 


\section*{Acknowledgements}

DG acknowledges support from the Core Research Grant CRG/2023/001448 of
the Anusandhan National Research Foundation (ANRF) of the Gov. of India. We thank Rajat Sharma and Farman Ullah for useful comments.


\bibliography{biblio}

\end{document}